\documentstyle[12pt,twoside,fleqn,epsfig,wrapfig,espcrc1]{article}

\newcommand{\AmS}{{\protect\the\textfont2
  A\kern-.1667em\lower.5ex\hbox{M}\kern-.125emS}}

\title{Highlights on Dilepton and Photon Observables}

\author{Itzhak Tserruya\address{Weizmann Institute of Science,\\
        Rehovot 76100, Israel }
        \thanks{e-mail: tserruya@ceres.weizmann.ac.il.
                 Work supported by the Israeli Science Foundation,
                 the MINERVA Foundation and the German-Israeli
                 Foundation for Scientific Research and Development}}

\begin{document}

\maketitle

\begin{abstract}

The highlights of Quark Matter 97 on dileptons and direct photons are presented.

\end{abstract}

\section{Introduction}
    Evidence of enhanced dilepton production  in S induced collisions
at 200 A GeV was first reported at the 95 Quark Matter Conference by the three
experiments involved in the measurement of dileptons at the CERN 
SPS --CERES, HELIOS-3 and NA38 \cite{it-qm95}.  The enhancement is particularly large 
at low masses
(m = 0.2 -- 1.0 GeV/c$^2$)  but it is also significant at intermediate masses 
(m = 1.5 -- 2.5 GeV/c$^2$). In contrast, there was no clear evidence of a signal 
in the search of direct photons. Since then, these results have been at the focus of 
attention, triggering a strong interest mainly stimulated by the possibility that 
the low-mass excess could reflect the onset of chiral symmetry restoration 
\cite{li-ko-brown}.
 Preliminary results with the Pb beam, presented at the 96 Quark Matter Conference
by the CERES and NA50 experiments \cite{tu-qm96,scomparin-qm96},  show also 
an enhanced production of dileptons 
confirming at least qualitatively the results with the S beam. For recent reviews, see 
refs. \cite{ad-qm96,it-hiroshima}. 
In this Conference we have seen a wealth of new and interesting results, both on the 
experimental and theoretical fronts, related to the measurement of dileptons and 
photons at the CERN SPS. In this paper, I discuss the highlights in the field
and I supplement what was presented in the talks with some comments and 
suggestions for future work.
 
\section{Low-mass Dileptons}
\subsection{Experimental Results}

\begin{figure}[t]
\vspace{-1.0cm}
\begin{center}
\epsfig{file=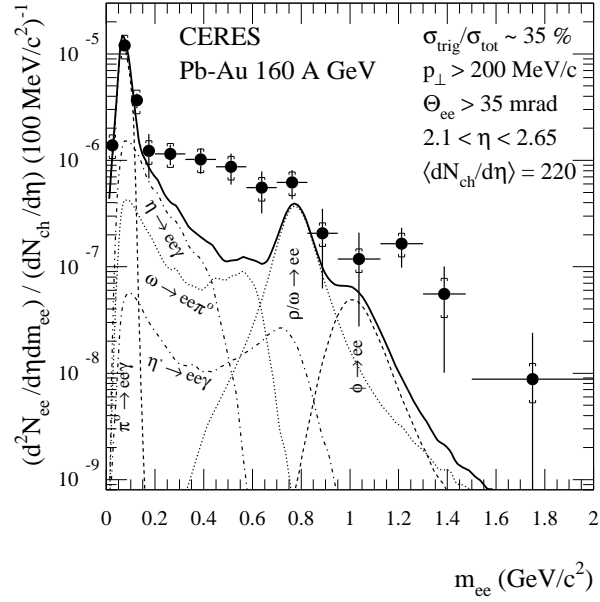,width=8.4cm}
\end{center}
\vspace{-1.5cm}
\caption{Inclusive $e^+e^-$ mass spectrum measured by CERES in 160 A~GeV Pb--Au 
         collisions normalized to the observed charged-particle density. The
         Figure also shows the summed and individual contributions from hadronic 
         sources ~\protect\cite{ir-qm97}.}
\label{fig:pbau95-mass}
\vspace{-0.5cm}
\end{figure}

   Ravinovich presented a detailed analysis of  low-mass electron pairs 
measured by CERES in Pb-Au collisions at 160 A GeV \cite{ir-qm97}. 
Fig. \ref{fig:pbau95-mass} shows the inclusive $e^+e^-$ mass spectrum 
normalized to the charged-particle rapidity density. The spectrum is very 
similar to the one measured in S-Au collisions \cite{prl95}. The 
yield is clearly enhanced compared to the predicted one on the basis of hadron decays 
shown by the thick solid line. 
The effect is most pronounced in the region from 300 to 700 MeV/c$^2$ where the 
enhancement factor is 5.8 $\pm$ 0.8 (stat) $\pm$ 1.5 (syst) but it also  
extends to higher masses. In the larger mass interval 0.2 $\le$ m $\le$ 
2.0 GeV/c$^2$ the enhancement factor is 3.5 $\pm$ 0.4 (stat) $\pm$ 0.9 (syst).

\begin{figure}[h!]
\vspace{-1.0cm}
\begin{center}
\epsfig{file=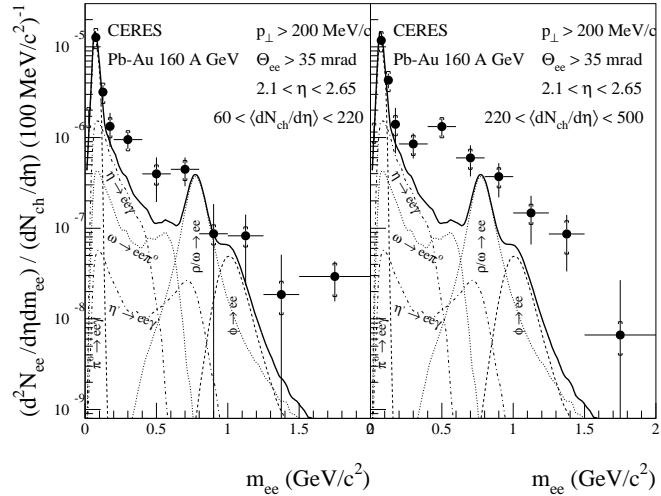,width=9cm}
\end{center}
\vspace{-1.0cm}
\caption{Inclusive $e^+e^-$ mass spectra measured by CERES in 160 A~GeV Pb--Au 
         collisions for low (left) and high (right) charged particle density.
         See also the caption of Fig. \ref{fig:pbau95-mass} ~\protect\cite{ir-qm97}.}
\label{fig:pbau95-two-mass}
\vspace{-0.5cm}
\end{figure}

Ravinovich presented also new results on  multiplicity and $p_t$ dependence. 
In spite of the limited statistics, the results show \cite{ir-qm97} that the yield, 
integrated over mass, increases faster than linearly with multiplicity. Another 
significant feature is that the shape of the excess also changes with multiplicity 
as shown in Fig. \ref{fig:pbau95-two-mass}. For masses m $>$ 200 MeV/c$^2$ 
the yield is enhanced over the entire $p_t$ spectra, but it appears more 
pronounced at low $p_t$.  This new information should be very valuable 
to further test and constrain the models put forward to explain the low-mass 
enhancement. 
 
\begin{figure}[t]
\vspace{-0.7cm}
\begin{center}
\epsfig{file=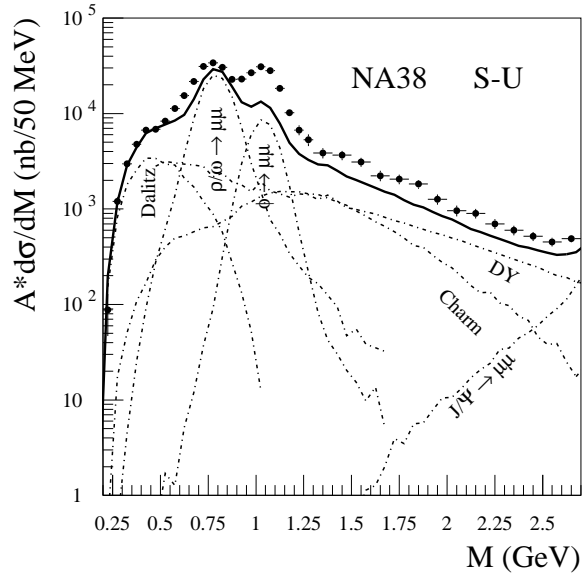,width=8cm}
\end{center}
\vspace{-0.5cm}
\caption{Inclusive $\mu^+\mu^-$ mass spectra measured by NA38 in 
              200 A~GeV S-U collisions. The thick line represents the summed 
              yield  of all known sources. The individual contributions
              are also shown ~\protect\cite{falco-qm97}.}
\label{fig:na38-su}
\end{figure}

   Falco presented very interesting results of a systematic study focussed 
on the emission of low-mass dimuon pairs performed by NA38 and including
p-U, S-S and S-U collisions at 200 A GeV \cite{falco-qm97}.  
   This is a most welcome ``new arrival'' since it is always good to have more than
one experiment looking at the same phenomena. Whereas the p-U data are well reproduced
by a cocktail of hadronic sources (with the somewhat uncertain extrapolation of the 
Drell-Yan 
contribution into low masses), the S data shows an enhancement of low-mass pairs. 
The enhancement is most apparent in the S-U collision system and there it clearly
extends over the intermediate mass region as illustrated in Fig.\ref{fig:na38-su}.
 
There is a striking difference in the shape of the low-mass dilepton spectrum as 
measured by CERES and by NA38. A pronounced  structure due to the resonance decays 
is clearly visible in the NA38 spectrum, whereas in the S and Pb CERES results the 
structure is completely washed out (see Figs.\ref{fig:pbau95-mass},
\ref{fig:pbau95-two-mass} and \ref{fig:sau+li+wambach+zahed}), raising the 
question of consistency between the two experiments. Resolution effects can 
be readily ruled out  since the low-mass spectrum in p-Be and p-Au collisions 
measured by CERES with the same apparatus clearly shows the $\rho$/$\omega$ peak 
\cite{pbe-paper}. We note also that the two experiments cover nearly symmetric ranges 
around mid-rapidity ($\eta$ = 2.1 -- 2.65 and $\eta$ = 3 -- 4 in CERES and NA38 
respectively). But there are also significant differences: CERES has a relatively 
low $p_t$ cut of 200 MeV/c on each track whereas NA38 is restricted to  
$m_t > $ 0.9 + 2($y_{lab} - 3.55)^2$ GeV/c$^2$. Moreover, and 
probably more significantly, NA38 has no centrality selection in the trigger 
whereas the CERES data corresponds to the top 35\% of the geometrical cross 
section. Given enough statistics it should be fairly easy for the two experiments 
to apply off-line common $m_t$ and centrality  cuts thereby  making possible a 
direct and meaningful comparison between their results.

\subsection{Interpretations}

\begin{figure}[t]
\vspace{-2.0cm}
\begin{center}
\epsfig{file=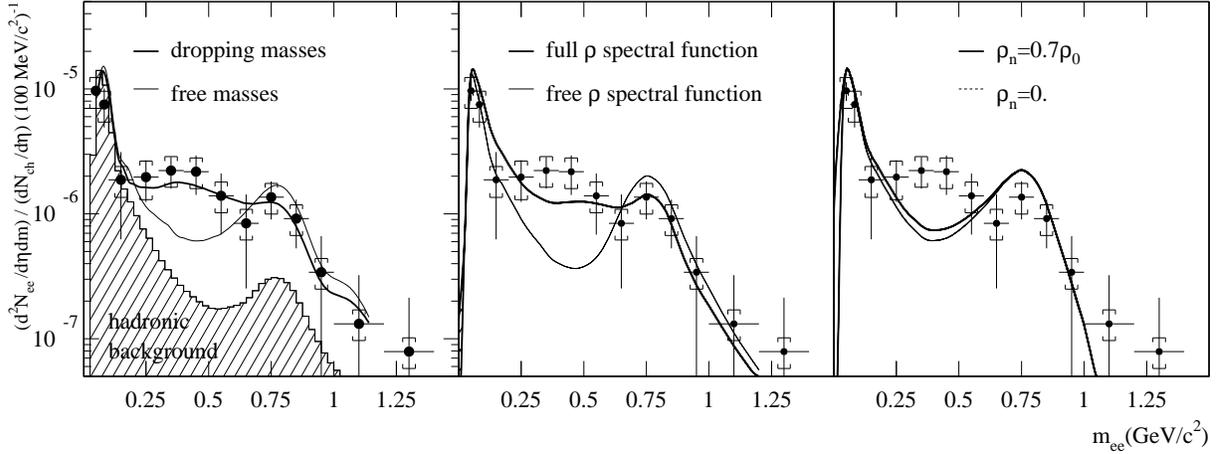,width=17cm}
\end{center}
\vspace{-1.3cm}
\caption{Inclusive $e^+e^-$ mass spectra measured by CERES in 200 A~GeV S--Au 
         collisions compared to calculations of Li, Ko and Brown
          ~\protect\cite{li-ko-brown} (left panel), Wambach 
          ~\protect\cite{wambach-qm97,rapp-wambach} (middle panel)  and Zahed 
          ~\protect\cite{zahed-qm97,zahed} (right panel).}
\label{fig:sau+li+wambach+zahed}
\vspace{-0.5cm}
\end{figure}

    The S results of CERES and HELIOS-3 on low-mass dileptons have triggered a
wealth of theoretical activity. A comprehensive theoretical review was given by 
Wambach \cite{wambach-qm97} and very interesting results were presented by Li  
\cite{li-qm97} and Zahed \cite{zahed-qm97}.
For an easier discussion and comparison between the different models I have plotted
in Fig. \ref{fig:sau+li+wambach+zahed} their results using the same scale 
and the same presentation of the S CERES data with statistical (vertical bars) and 
systematic (brackets) errors plotted independently.
The histogram in the left panel represents the summed contribution from hadronic sources
assuming a simple superposition of pp collisions. There is a consensus that an 
additional source beyond this simple assumption is needed. Furthermore, it is commonly 
recognized that the pion annihilation channel ($\pi^+\pi^- \rightarrow l^+l^-$), 
obviously not present in pp collisions, has to be taken into account. This channel 
has been included in all recent calculations and although 
it accounts for a large fraction of the 
observed enhancement (the  thin line in the three panels contain the pion annihilation 
in addition to the hadron decays) it is not sufficient and the 
calculations  fail to reproduce the data in the mass region  
0.2 $< m_{e^+e^-} <$ 0.5 GeV/c$^2$. These data have been quantitatively 
explained by taking into account in-medium modifications of the vector mesons.
Li, Ko and Brown \cite{li-ko-brown} were the first to propose and use a decrease of 
the $\rho$-meson mass in the hot and dense fireball as a precursor of chiral 
symmetry restoration. With this approach,  an excellent agreement was achieved  
\cite{li-ko-brown,cassing} with the CERES data as demonstrated by the solid line 
in the left panel of Fig. \ref{fig:sau+li+wambach+zahed}. 

 Wambach discussed in detail the results of a different approach, using a $\rho$-meson 
spectral function which takes into account the $\rho$ propagation in hot and dense
matter, 
including in particular the pion modification in the nuclear medium and 
the scattering of $\rho$ mesons off baryons. This leads to a large 
broadening of the $\rho$-meson line shape and consequently to a considerable 
enhancement of low-mass dileptons. These calculations also reproduce  the CERES  
S data as can be seen in the middle panel of Fig. \ref{fig:sau+li+wambach+zahed}.

One should also notice that these two approaches reproduce equally well the 
S-W results of HELIOS-3 and the preliminary CERES results of Pb-Au 
collisions \cite{ko-private,rapp-wambach,cassing-wambach}. 
Their underlying physical picture is, however, very different. One case
invokes chiral symmetry restoration  which is an intrinsic property, 
whereas the second uses collision broadening which is a dynamic effect. 
More stringent tests are needed to help discriminate between the two approaches.
The new results on  multiplicity and $p_t$ dependences together with forthcoming 
results with reduced  errors should be very helpful. Furthermore, the calculations 
themselves are not free of debate. In this context the 
results  presented by Zahed at this Conference \cite{zahed-qm97} are very
interesting. 
His approach is in principle similar to the one of Wambach. He addresses the same physics 
of in-medium modifications of the $\rho$ spectral function, using on-shell chiral reduction
formulas and enforcing known constraints,  and yet, with a realistic nucleon density
$\rho_N$,  he fails to reproduce the low-mass data as shown in the right panel of 
Fig. \ref{fig:sau+li+wambach+zahed} \cite{zahed-qm97,zahed}.

\section{Vector Mesons}

\begin{figure}[t]
\vspace{-1.5cm}
\begin{center}
\epsfig{file=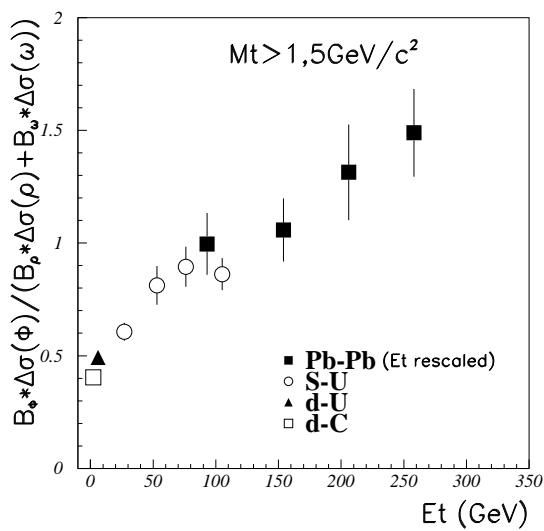,width=8.0cm}
\end{center}
\vspace{-1.5cm}
\caption{The $\phi$/($\rho + \omega$) ratio for various collision systems
         as a function of E$_t$ measured by NA38/50 ~\protect\cite{jouan-qm97}.}
\label{fig:jouan-phiom}
\vspace{-0.5cm}
\end{figure}

    Jouan \cite{jouan-qm97} presented new results from NA50 on the 
production of the vector mesons $\rho,\omega$ and $\phi$ in Pb-Pb collisions.
There is now an impressive set of data covering a variety of systems
d-C, d-U, S-U and Pb-Pb collisions. The ratio $\phi$/($\rho+\omega)$
for all these systems is shown in Fig. \ref{fig:jouan-phiom} as a function of 
$E_t$. The ratio increases approximately linearly with centrality and also with
the size of the system, by almost a factor of four from d-C to Pb-Pb collisions.
Historically, the data on vector meson production as measured by HELIOS and NA38
has always been presented in the form of the ratio $\phi$/($\rho+\omega)$.
The increase of the ratio was taken as a measure of the increase of the 
$\phi$-meson production --which is a very interesting feature in the context 
of strangeness 
enhancement-- with the implicit assumption that the $\omega$ and $\rho$
meson production remains unaffected. 
However, since the $\rho$ meson is suspected to change its properties, 
as discussed in the previous section (either the width, or the mass or both), the 
amount of $\phi$ enhancement cannot be readily assessed. For that it would 
be more appropriate to see the yield of the $\phi$ and ($\rho+\omega$) mesons 
plotted separately.
 
\section{Intermediate-Mass Dileptons}
   We have not seen new experimental results on dilepton production at intermediate
masses 1.5 $<$ m $<$ 3.0 GeV/c$^2$ at this Conference. But we did see new calculations
presented by Li \cite{li-qm97} in comparison with the HELIOS-3 results on 
central S-W collisions at 200 A GeV. He considers on top of the {\it physics}
background of Drell-Yan and open charm pairs, the thermal radiation of muon pairs
resulting from secondary meson interactions including higher resonances and in particular 
the $\pi a_1 \rightarrow l^+l^-$. His calculations are based on the same relativistic 
\begin{figure}[t]
\vspace{-1.0cm}
\begin{minipage}[t]{75mm}
\epsfig{file=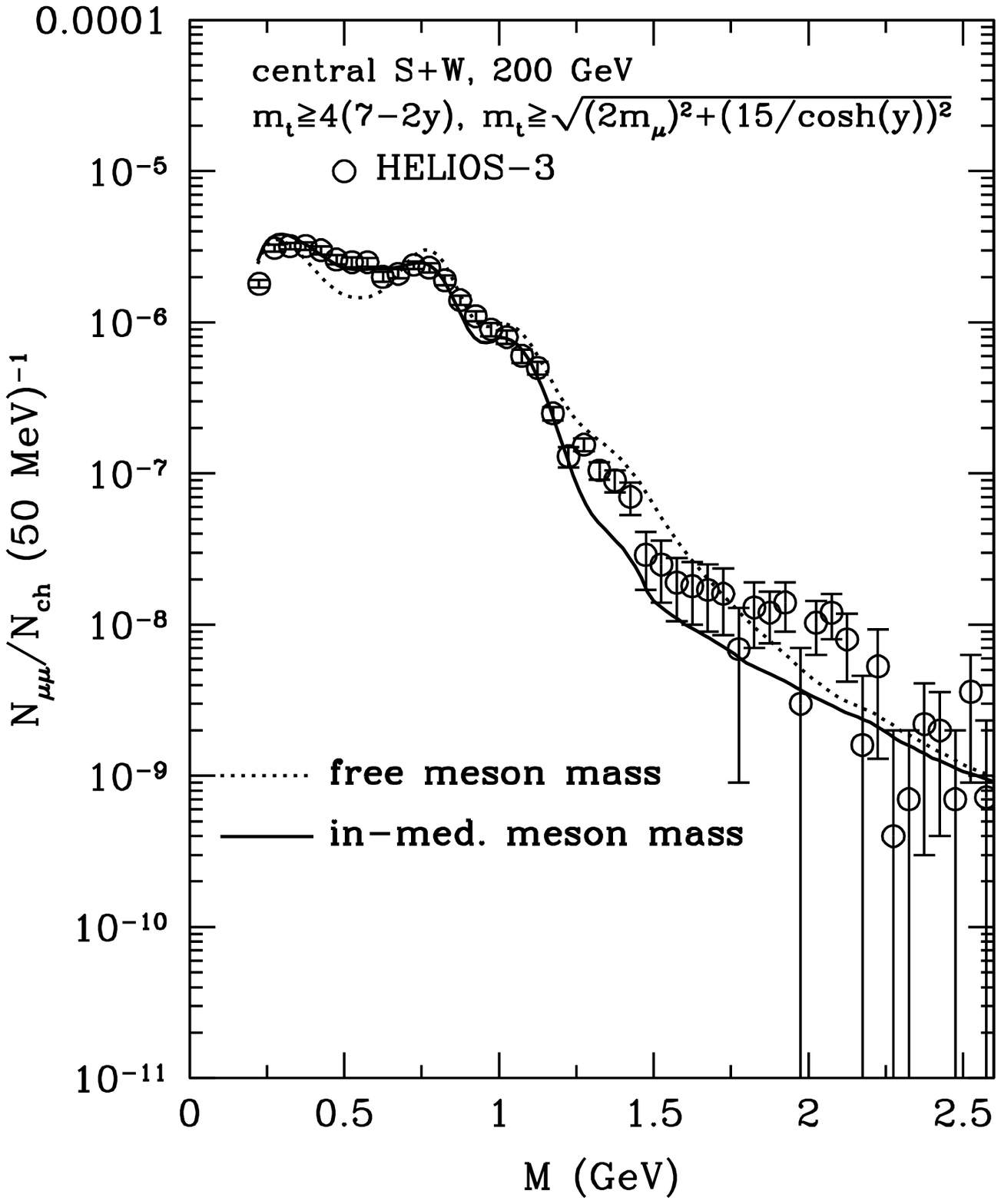,width=7.5cm,
height=7.0cm}
\vspace{-1.1cm}
\caption{Calculations of Li with free and in-medium meson masees compared to the 
         HELIOS-3 dimuon data  ~\protect\cite{li-qm97}.}
\label{fig:int-mass-li1}
\end{minipage}
\hspace{\fill}
\begin{minipage}[ht]{75mm}
\vspace{-4.8cm}
\epsfig{file=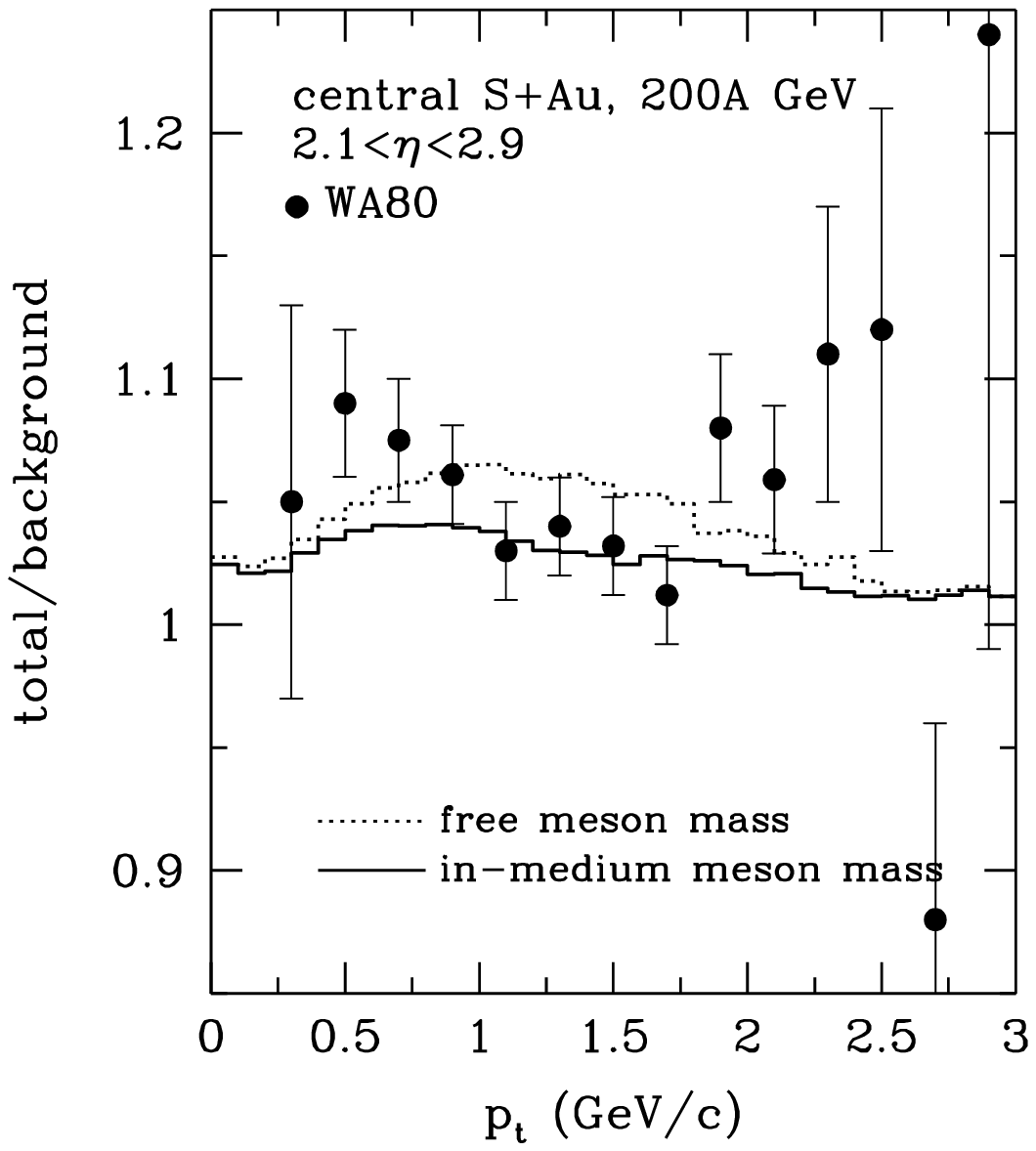,height=7.4cm,
width=7.5cm}
\vspace{-1.5cm}
\caption{Comparison of the calculations of Li with the WA80 photon data in S-Au 
         collisions at 200 A GeV ~\protect\cite{li-qm97}.}
\label{fig:li-wa80photons}
\end{minipage}
\vspace{-0.5cm}
\end{figure}
transport model used to calculate the low-mass dileptons discussed in the previous 
section. His results are presented in Fig. \ref{fig:int-mass-li1} showing
the total yield (physics background + thermal yield) with the assumption
of  free masses (dotted line) and dropping vector meson masses (solid line).
The latter leads to a much better agreement with the data at low
masses (from 0.3 to 0.7 GeV/c$^2$), as already mentioned in the previous section,
whereas in the intermediate mass region the difference between free and in-medium
meson masses with respect to the data is not so large. The calculations with
free masses slightly overestimate the data whereas with dropping masses 
the situation is reversed.  The intermediate mass region alone cannot justify the use
of a dropping mass model, however it is important that this model can explain
simultaneously the low and intermediate mass regions.

\section{Photons}
 In contrast with the dilepton results, there is no clear evidence of enhancement
in the measurements of real photons. All experiments performed to date with O and S 
beams have been able to establish only an upper limit for the production of 
thermal photons, which is of the order of 10 -- 15\%  of the expected yield 
from hadron decays. The sensitivity is actually limited not by statistics but by
the systematic errors. 
At this conference, WA98 has presented preliminary results on the search of 
direct photons in Pb-Pb collisions which show a slightly larger effect, an 
excess of $\sim$20\% over the hadronic background.

   Direct photons are expected  to provide analogous information 
to thermal dileptons since real and virtual photons should carry the same physics
information. Therefore, a simultaneous quantitative  description of results 
on low-mass dileptons and direct photons  within a single model is a significant  
step in establishing a consistent and reliable interpretation 
of experimental results.
  Li has shown a first attempt in this direction \cite{li-brown}.  
With the same fireball model including dropping masses used to explain the 
CERES and HELIOS-3 dilepton results he calculated the amount of direct photons
in central S-Au collisions at 200 A GeV. The results are presented  in 
Fig. \ref{fig:li-wa80photons} in the form of the ratio of the total photon yield to 
the hadronic background from $\pi^0$ and $\eta$ decays. The excess of 
direct photons is predicted to be a few
percent of the hadronic background, in agreement with the experimental results 
and with a simple estimate based on order of magnitude considerations \cite{it-qm95}. 
This sharpens the strict requirement imposed on the experiments to control the 
systematic errors down to the percent level in order to be able to observe direct photons.

\vspace{-0.3cm}
\section*{Acknowledgements}
   It is a pleasure to thank all the speakers of the dilepton plenary and parallel 
sessions, and in particular A. Falco, D. Jouan, G.Q. Li, I. Ravinovich, J. Wambach
and I. Zahed, for providing the material I needed for my oral and written
presentations.

\end{document}